\documentclass[10pt,a4paper]{iopart}
\usepackage{graphicx}

\begin{document}

\title[Electron and phonon dispersions of the 2D Holstein model]{Electron and phonon dispersions of the two dimensional Holstein model: Effects of vertex and non-local corrections}

\author{J.P.Hague}
\address{Max-Planck Institut f\"{u}r Phyisk Komplexer Systeme, Dresden, Germany}
\date{22nd April 2003}

\begin{abstract}
I apply the newly developed dynamical cluster approximation (DCA) to
the calculation of the electron and phonon dispersions in the two
dimensional Holstein model. In contrast to previous work, the DCA
enables the effects of spatial fluctuations (non-local corrections) to
be examined. Approximations neglecting and incorporating lowest-order
vertex corrections are investigated. I calculate the phonon density of
states, the renormalised phonon dispersion, the electron dispersion
and electron spectral functions. I demonstrate how vertex corrections
stabilise the solution, stopping a catastrophic softening of the
$(\pi,\pi)$ phonon mode. A kink in the electron dispersion is found in
the normal state along the $(\zeta,\zeta)$ symmetry direction in both
the vertex- and non-vertex-corrected theories for low phonon
frequencies, corresponding directly to the renormalised phonon
frequency at the $(\pi,0)$ point. This kink is accompanied by a sudden
drop in the quasi-particle lifetime. Vertex and non-local corrections
enhance the effects at large bare phonon frequencies. [PUBLISHED AS
J.PHYS.:CONDENS.MATTER, {\bf 15} (2003) 2535-2550 ON 22ND APRIL 2003] 
\end{abstract}

\section{Introduction}
\label{section:introduction}

Interest in the electron-phonon problem has recently been rekindled by
the discovery of large electron-phonon couplings in the cuprate
superconductors and the CMR manganites
\cite{zhao1996a,lanzara2001a,mcqueeny1999a}. Strong electron-phonon
interactions outside the regime of traditional theories are also
thought to be important in the bismuthates and in potassium doped
Buckminsterfullerene superconductors \cite{freericks1998a}.

One of the best known approaches to the electron-phonon problem is
that of Migdal and Eliashberg
\cite{migdal1958a,eliashberg1960a}. Migdal showed that the effects of
vertex corrections and momentum on the self-energy should be
negligible if the phonon frequency is much less than the Fermi energy
\cite{migdal1958a}. Migdal's method was developed further by
Eliashberg to investigate the superconducting state
\cite{eliashberg1960a}. The physical content of Migdal's theorem is
that there is a high probability that the most recently emitted phonon
is the first phonon to be reabsorbed. This is true, provided excited
states remain close to the Fermi-energy, i.e. phonon energies and
couplings are small compared with the band width. Within Migdal
theory, it is therefore acceptable to neglect processes in which the
order of emission and reabsorption is changed. When these techniques
were introduced, no materials had been found where the electron-phonon
coupling was strong, and the phonon frequency large compared to the
Fermi energy. For this reason, the application of Migdal--Eliashberg
(ME) theory was very successful, and remains highly
regarded. Unfortunately, strong electron-phonon coupling and large
phonon frequencies in the systems detailed in the first paragraph is
likely to make them incompatible with the Migdal--Eliashberg approach,
and further extensions to the theory seem necessary.  The aim of this
paper is to evaluate and discuss the effects of both vertex
corrections and non-local fluctuations due to the DCA on the theory of
coupled electron-phonon systems.

Previous attempts to extend ME theory include the introduction of
vertex corrections into the Eliashberg equations by Grabowski and Sham
\cite{grabowski1984a}, and an expansion to higher order in the Migdal
parameter by Kostur and Mitrovi\'{c} in order to investigate the 2D
electron-phonon problem \cite{kostur1993a}. Grimaldi \emph{et al.}
generalised the Eliashberg equations to include momentum dependence
and vertex corrections \cite{grimaldi1995a}. A discussion of the
applicability of these and other approximations to the vertex function
can be found in reference \cite{danylenko2001a}.

The importance of vertex corrections to electron-phonon problems
within the local or dynamical mean-field approximation (DMFA), has
been investigated by Miller \emph{et al.} \cite{miller1998a} and
Deppeler \emph{et al.} \cite{deppeler2002a}. In these studies, the
effects of both electrons on phonons and phonons on electrons are
treated with equal importance, leading to a fully self-consistent
theory.  Miller \emph{et al.} focus on situations where the effects of
vertex corrections can be handled using a Coulomb pseudopotential
extension to the basic ME theory, and where the explicit treatment of
the vertex is essential. Deppeler \emph{et al.} take the expansion to
third order in the self-energy. The authors calculate physical
observables, such as isotope coefficients, phonon spectral functions
and superconducting transition temperatures.

The current paper goes beyond the previous work by introducing a fully
self-consistent momentum-dependent self-energy to the problem via the
dynamical-cluster approximation (DCA). DCA is a new technique, which
extends the DMFA by introducing short-range fluctuations in a
controlled manner \cite{hettler1998a}. I apply the DCA to the
calculation of electron and phonon dispersion curves in the Holstein
model, which is one of the simplest non-trivial models of
electron-phonon systems. Two approximations for the electron and
phonon self energies are applied. The first neglects vertex
corrections, but incorporates non-local fluctuations. The second
incorporates lowest order vertex and non-local corrections. The vertex
corrections allow the sequence of phonon absorption and emission to be
reordered once, and therefore introduce exchange effects. At each
stage, the DCA result is compared to the corresponding DMFA result. In
principal, with parameters close to the Migdal regime, this
enhancement of ME theory should be sufficient to correct the results,
and the series allowing for additional reorderings should be
convergent. However, such an approach may not work when the phonon
frequency greatly exceeds the Fermi-energy, known as the instantaneous
limit.

This paper is organised as follows. In section \ref{section:dca}, the
DCA is introduced. In section \ref{section:holstein} the Holstein
model of electron-phonon interactions is described, and the
perturbation theory and the full algorithm used in this work are
detailed. In section \ref{section:results} the results are
presented. They are divided into two parts. First the effects of
non-local and vertex corrections on the phonon density of states and
renormalised phonon frequency are discussed. Then the electron
dispersion and spectral function are computed. A summary of the major
findings of this research is given in section \ref{section:summary}.

\section{The dynamical cluster approximation}
\label{section:dca}

\begin{figure}[t]
\begin{indented}\item[]
\resizebox{60mm}{!}{\includegraphics{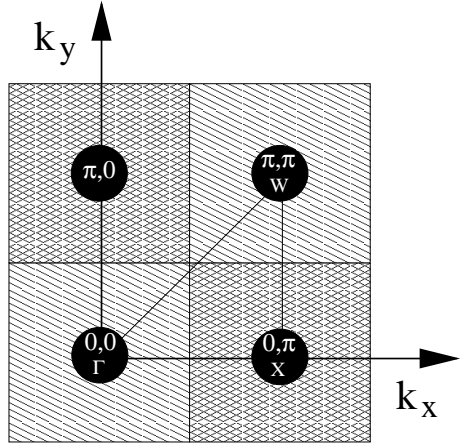}}
\end{indented}
\caption{A schematic representation of the reciprocal-space coarse
graining scheme for a 4 site DCA. Within the shaded areas, the
self-energy is assumed to be constant. There is a many to one mapping
from the crosshatched areas to the points at the centre of those
areas. The coarse graining procedure corresponds to the mapping to a
periodic cluster in real space, with spatial extent
$N_{c}^{1/D}$. Also shown are the high symmetry points $\Gamma$, $W$
and $X$, and lines connecting the high symmetry points. Since an
infinite number of $\mathbf{k}$ states are involved in the many to one
mapping, the approximation is in the thermodynamic limit. The DMFA
corresponds to $N_{C}=1$.}
\label{fig:coarsegrain}
\end{figure}

\begin{figure}[t]
\begin{indented}\item[]
\resizebox{80mm}{!}{\includegraphics{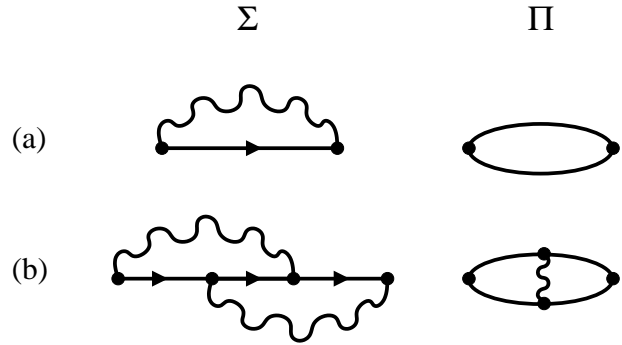}}
\end{indented}
\caption{Diagrammatic representation of the current
approximation. Series (a) represents the vertex-neglected theory which
corresponds to the Migdal--Eliashberg approach. This is valid when
there is a high probability that the last emitted phonon is the first
to be reabsorbed, which is true if the phonon energy $\omega_0$ and
electron-phonon coupling $U$ are small compared to the Fermi
energy. Series (b) represents additional diagrams for the vertex
corrected theory. The inclusion of the lowest order vertex correction
allows the order of absorption and emission of phonons to be swapped
once. In principal, for moderate phonon frequency and electron-phonon
coupling, these additions to the theory should be sufficient to
correct the results of the vertex-neglected theory. The phonon self
energies are labelled with $\Pi$, and $\Sigma$ denotes the electron
self-energies. Lines represent the full electron Green's function and
wavy lines the full phonon Green's function.}
\label{fig:feynmandiag}
\end{figure}

The dynamical cluster approximation \cite{hettler1998a,hettler2000a} is a
recent extension to the dynamical mean-field approximation. DMFA has been
extensively applied to solving infinite-dimensional lattice models, where the
DMFA formalism becomes exact \cite{metzner1989a, georges1996a}. It has also
been used to approximate two and three dimensional systems, where it is known
as the local approximation \cite{tahvildar1997a,miller1998a}. In DMFA, the
self-energy is assumed to be local, and $\Sigma(\mathbf{k},z)$ is replaced by
its momentum-independent counterpart, $\Sigma(z)$. As such, the Green's
function may be coarse grained across the Brillouin zone,
\begin{equation}
\label{eqn:greensfunction}
G(z)=\int _{-\infty }^{\infty }\frac{{\mathcal{D}}(\epsilon )\, d\epsilon }{z+\mu-\epsilon -\Sigma (z)}
\end{equation}
where \( {\mathcal{D}}(\epsilon ) \) is the non-interacting Fermion
density of states (DOS). The chemical potential is represented by
$\mu$, and $z$ may take the values $i\omega_n$ or $\omega+i\eta$
depending on whether one wants to work with the Matsubara or real axis
Green's function ($\eta$ is a small value).

Application of the DMFA to one and two dimensional models is expected
to give an incomplete description of the physics. The DMFA can be
thought of as the first term of an expansion in $1/n_{\mathrm{nn}}$,
where $n_{\mathrm{nn}}$ is the coordination number
\cite{georges1996a}. For three dimensional systems, $n_{\mathrm{nn}}$
ranges between 6 and 12, and it is often argued that additional terms
give a minor correction. In contrasts additional terms in the
$1/n_{\mathrm{nn}}$ expansion are expected to be important in 1 and 2
dimensions. An example of significant differences between the one, two
and three dimensional cases comes from isotropic-exchange
spin-systems. In 3D, a finite transition temperature is found,
consistent with the mean-field result (although the critical exponent
is modified). However, significant non-local fluctuations in one and
two dimensions are expected to reduce the N\'{e}el temperature to zero
(Mermin--Wagner theorem), and the mean-field approach fails
completely.

The DCA follows a similar procedure to the DMFA. The Brillouin zone is
divided up into $N_C$ subzones consistent with the lattice symmetry
(see figure \ref{fig:coarsegrain}). Within each of these zones, the
self-energy is assumed to be momentum independent. This leads to the
relation,
\begin{equation}
\label{eqn:dcagreensfunction}
G(\mathbf{K}_{i},z)=\int _{-\infty }^{\infty }\frac{{\mathcal{D}}_{i}(\epsilon )\, d\epsilon }{z+\mu-\epsilon -\Sigma (\mathbf{K}_{i},z)}
\end{equation}
where \( {\mathcal{D}}_{i}(\epsilon ) \) is the non-interacting
Fermion density of states for subzone $i$. The $\mathbf{K}_i$ are the
average $\mathbf{k}$ for each subzone, plotted as the large dots in
figure \ref{fig:coarsegrain}. This partial treatment of momentum
dependence in the theory introduces non-local fluctuations with a
characteristic length scale of $N_C^{1/D}$.

The combination of the self-energy and the coarse grained Green's
function leads to a modified Dyson equation,
\begin{equation}
\label{eqn:moddyson}
{\mathcal{G}}_{0}^{-1}(\mathbf{K}_{i},z)-G^{-1}(\mathbf{K}_{i},z)=\Sigma (\mathbf{K}_{i},z)
\end{equation}
The self-consistent condition is closed by calculating the self-energy
from ${\mathcal{G}}_{0}(z,\mathbf{K})$, which has the interpretation
of the host Green's function of a cluster impurity model (sometimes
known as the cluster excluded Green's function). Different
approximations exist for the calculation of the self-energy, including
quantum Monte-Carlo and perturbative approaches.

Both dynamical-mean-field and dynamical-cluster approximations are
calculated in the thermodynamic limit, as opposed to conventional
finite sized techniques where the particle number is equivalent to the
cluster size. In this work, the DCA is used, since it involves only a
small amount of additional computing time per iteration, while rapid
convergence in cluster size is expected. Estimates based on the
convergence of the Hubbard problem (which has a relatively strong
momentum dependence) suggest that the Hubbard model is well
approximated by a cluster size of $N_C=64$ \cite{jarrell2001a}.

There are a number of methods that can be applied to speed up the DCA
algorithm. Firstly, the symmetry of the problem may be taken into
account through the appropriate \emph{planar} point group
$\mathrm{p}m3m$ \cite{internationaltables}. In the large cluster
limit, this means that only $1/8$th of the total momentum points need
to be used in the calculation. Similar considerations can be used to
dramatically reduce the number of terms in the momentum summations
needed to calculate the self-energy terms. This is particularly
important when vertex corrections are included, since the calculation
scales as $N^3$. The Matsubara frequency sums are, in principle,
infinite. However, in this work they were truncated at a frequency
where asymptotic behaviour is obeyed to a certain accuracy. All points
above this frequency which were needed in the calculation of another
quantity were computed from the asymptotic function, ensuring an
accurate self-consistent procedure. The partial DOS are central to the
DCA and were calculated using the analytic tetrahedron method to
ensure very high accuracy \cite{lambin1984a}. All code used in this
study was written in object oriented c++ to make the implementation as
simple and reliable as possible.

\section{The Holstein model}
\label{section:holstein}

A simple, yet non-trivial, model of electron-phonon interactions
treats phonons as individual nuclei vibrating in a static harmonic
potential (representing the interaction between all nuclei) i.e. only
one frequency $\omega_0$ is considered. The phonons couple to the
local electron density via a simple momentum-independent coupling
constant $g$. The resulting \emph{Holstein Hamiltonian}
\cite{holstein1959} is written as,
\begin{equation}
H=-\sum_{ij\sigma}t_{<ij>\sigma}c^{\dagger}_{i\sigma}c_{j\sigma}+\sum_{i\sigma} n_{i\sigma} (gr_i-\mu)+\sum_i\left( \frac{M\omega_{0}^2r_i^2}{2}+\frac{p_i^2}{2M}\right)
\end{equation}
The first term in this Hamiltonian represents a tight binding model with
hopping parameter $t$. Its Fourier transform takes the form
$\epsilon_{k}=-2t\sum_{i=1}^{D}\cos(k_{i})$. The second term connects the
local ion displacement, \( x_{i} \) to the local electron density. Finally the
last term can be identified as the bare phonon Hamiltonian. The creation and
annihilation of electrons is represented by \( c^{\dagger }_{i} \)(\( c_{i}
\)), \( p_{i} \) is the ion momentum and \( M \) the ion mass

It is possible to find an expression for the effective interaction between
electrons by integrating out phonon degrees of freedom \cite{bickers1989}. In
Matsubara space, this interaction is:
\begin{equation}
\label{eqn:phononfn}
U(i\omega _{s})=\frac{U\omega_{0}^{2}}{\omega _{s}^{2}+\omega _{0}^{2}}
\end{equation}
Here, $\omega _{s}=2\pi sT$ are the Matsubara frequencies for bosons and $s$
is an integer. A variable $U=-g^2/M\omega_0^2$ is defined to represent the
magnitude of the effective electron-electron coupling in the remainder of this
paper.

Interaction (\ref{eqn:phononfn}) has two important limits. When
$\omega_0\rightarrow 0$, the interaction is defined by a Kronecker
$\delta$ function. The DMFA of this limit was written down by
Freericks \emph{et al}, and solved by Millis \emph{et al.}
\cite{freericks1998a,millis1996a}. When phonon frequency and coupling
are small, Migdal's theorem applies. Migdal's approach allows vertex
corrections to be neglected, and results in a momentum-independent
theory corresponding to a local or dynamical mean-field solution. The
theorem is exact when $U=0^{-}$, $\omega_0=0^+$.

In the limit that $\omega_0\rightarrow\infty$, the Holstein model maps onto an
attractive Hubbard model. A particular feature of this mapping is that the
second order direct and exchange diagrams of the expansion in the bare
propagator develop the same functional form, and are of the same order of
magnitude, while the first-order diagram becomes constant and can be absorbed
into the chemical potential. Since the exchange diagram includes one vertex
correction, it is clear that vertex corrections are extremely important in
this limit. The vertex-corrected theory described in this paper has the
appropriate weak coupling behaviour for large $\omega_0$, unlike the
vertex-neglected theory.

In this paper, the perturbation theory shown in figure \ref{fig:feynmandiag}
is used, following directly from the free energy expansion of Baym and
Kadanoff \cite{miller1998a}. The electron self-energy has two terms,
$\Sigma_{\mathrm{ME}}(\omega,\mathbf{K})$ neglects vertex corrections (figure
\ref{fig:feynmandiag}(a)), and $\Sigma_{\mathrm{VC}}(\omega,\mathbf{K})$
corresponds to the vertex corrected case (figure
\ref{fig:feynmandiag}(b)). $\Pi_{\mathrm{ME}}(\omega,\mathbf{K})$ and
$\Pi_{\mathrm{VC}}(\omega,\mathbf{K})$ correspond to the equivalent phonon
self energies. The diagrams translate as follows (note that in the following 4
equations, $\mathbf{Q}\equiv (i\omega_n,\mathbf{Q})$)

\begin{equation}
\Sigma_{\mathrm{ME}}(\mathbf{K})=UT\sum_{\mathbf{Q}}G(\mathbf{Q})D(\mathbf{K}-\mathbf{Q})
\end{equation}
\begin{equation}
\Pi_{\mathrm{ME}}(\mathbf{K})=-2UT\sum_{\mathbf{Q}}G(\mathbf{Q})G(\mathbf{K}+\mathbf{Q})
\end{equation}
\begin{equation}
\Sigma_{\mathrm{VC}}(\mathbf{K})=(UT)^2\sum_{\mathbf{Q}_1,\mathbf{Q}_2}G(\mathbf{Q}_1)G(\mathbf{Q}_2)G(\mathbf{K}-\mathbf{Q}_1-\mathbf{Q}_2)D(\mathbf{K}-\mathbf{Q}_1)D(\mathbf{Q}_2-\mathbf{Q}_1)
\end{equation}
\begin{equation}
\Pi_{\mathrm{VC}}(\mathbf{K})=-(UT)^2\sum_{\mathbf{Q}_1,\mathbf{Q}_2}G(\mathbf{K})G(\mathbf{Q}_1)G(\mathbf{Q}_2)G(\mathbf{K}-\mathbf{Q}_1+\mathbf{Q}_2)D(\mathbf{Q}_2-\mathbf{Q}_1)
\end{equation}

The phonon propagator $D(z,\mathbf{K})$ is calculated from,
\begin{equation}
D(i\omega_s,\mathbf{K})=\frac{\omega_0^2}{\omega_s^2+\omega_0^2-\Pi(i\omega_s,\mathbf{K})}
\label{eqn:phonprop}
\end{equation}
and the Green's function from equation (\ref{eqn:dcagreensfunction})
$\Sigma=\Sigma_{\mathrm{ME}}+\Sigma_{\mathrm{VC}}$. It is possible to
extend equation \ref{eqn:phonprop} to deal with a momentum dependent
bare phonon dispersion, using a generalisation of the DMFA method in
\cite{motome2000a}. When extending the formalism to investigate a
momentum-dependent interaction term, the interaction term should be
coarse grained as in \cite{hettler2000a}. All the ingredients now
exist for the self-consistent loop. The algorithm is started with
$\Sigma,\Pi=0$, then the DCA propagators, $G$ and $D$ are calculated
from equations (\ref{eqn:phonprop}) and
(\ref{eqn:dcagreensfunction}). $\Pi$ is then calculated followed by
$D$ and $\Sigma$. The loop is completed by calculating $G$.

The double three-fold integration over momentum is the main barrier to
performing vertex-corrected calculations, meaning that only relatively small
clusters can be treated. Since the traditional phonon problem with $\omega_0,U
\ll W$ ($W$ is the bandwidth) has fluctuations which are almost momentum
independent \cite{migdal1958a}, it is expected that the DCA will have
especially fast convergence in $N_C$ for the parameter regime where
$\omega_0,U< W$, and that calculations with relatively small cluster size
accurately reflect the physics. In this respect, finite size calculations are
likely to fail, and the application of DCA to this problem is essential.

\section{Results}
\label{section:results}

\subsection{Phonon dispersion and DOS}
\label{sec:phonondispersion}

\begin{figure}[t]
\begin{indented}\item[]
\rotatebox{270}{\resizebox{!}{75mm}{\includegraphics{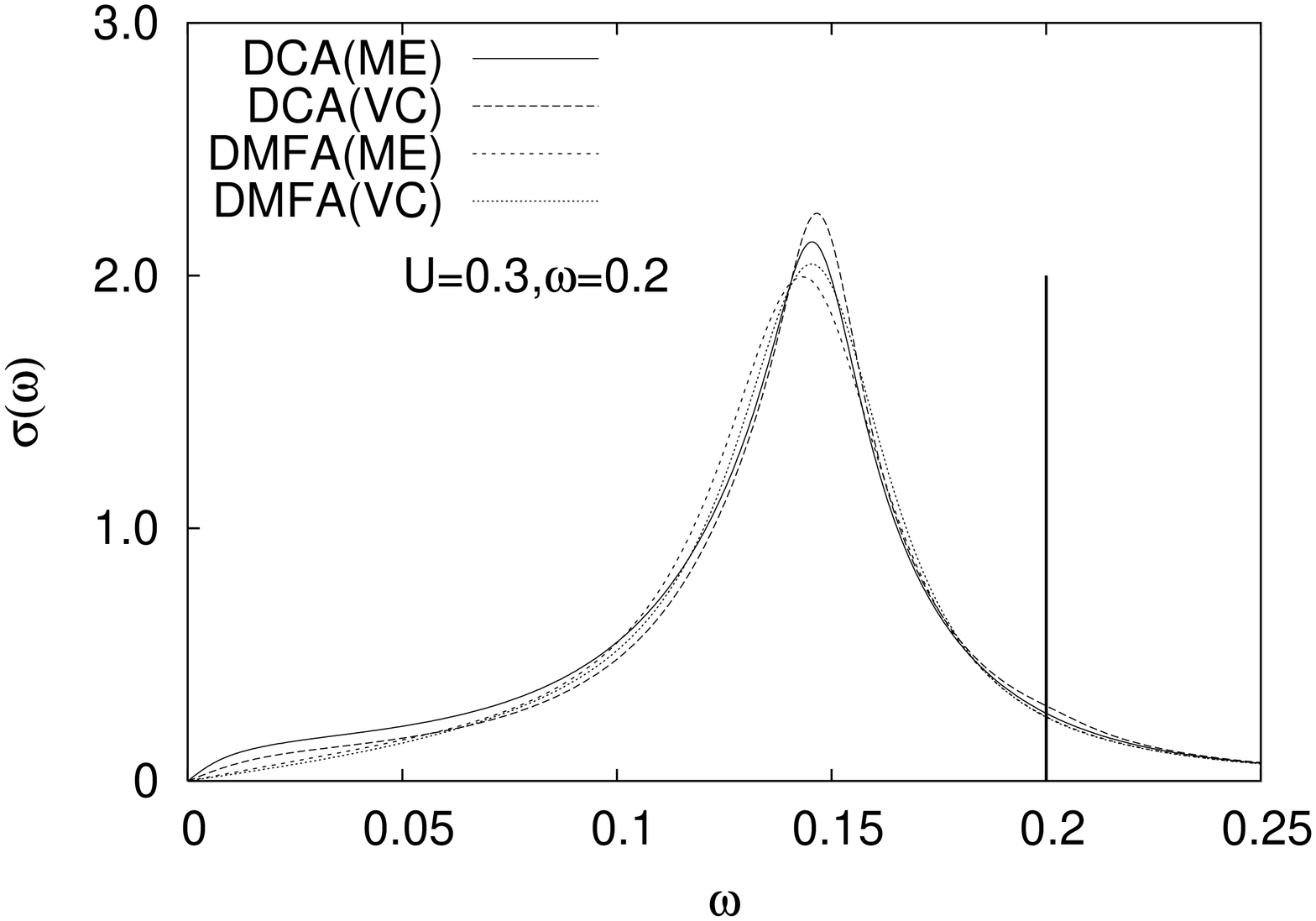}}}
\rotatebox{270}{\resizebox{!}{75mm}{\includegraphics{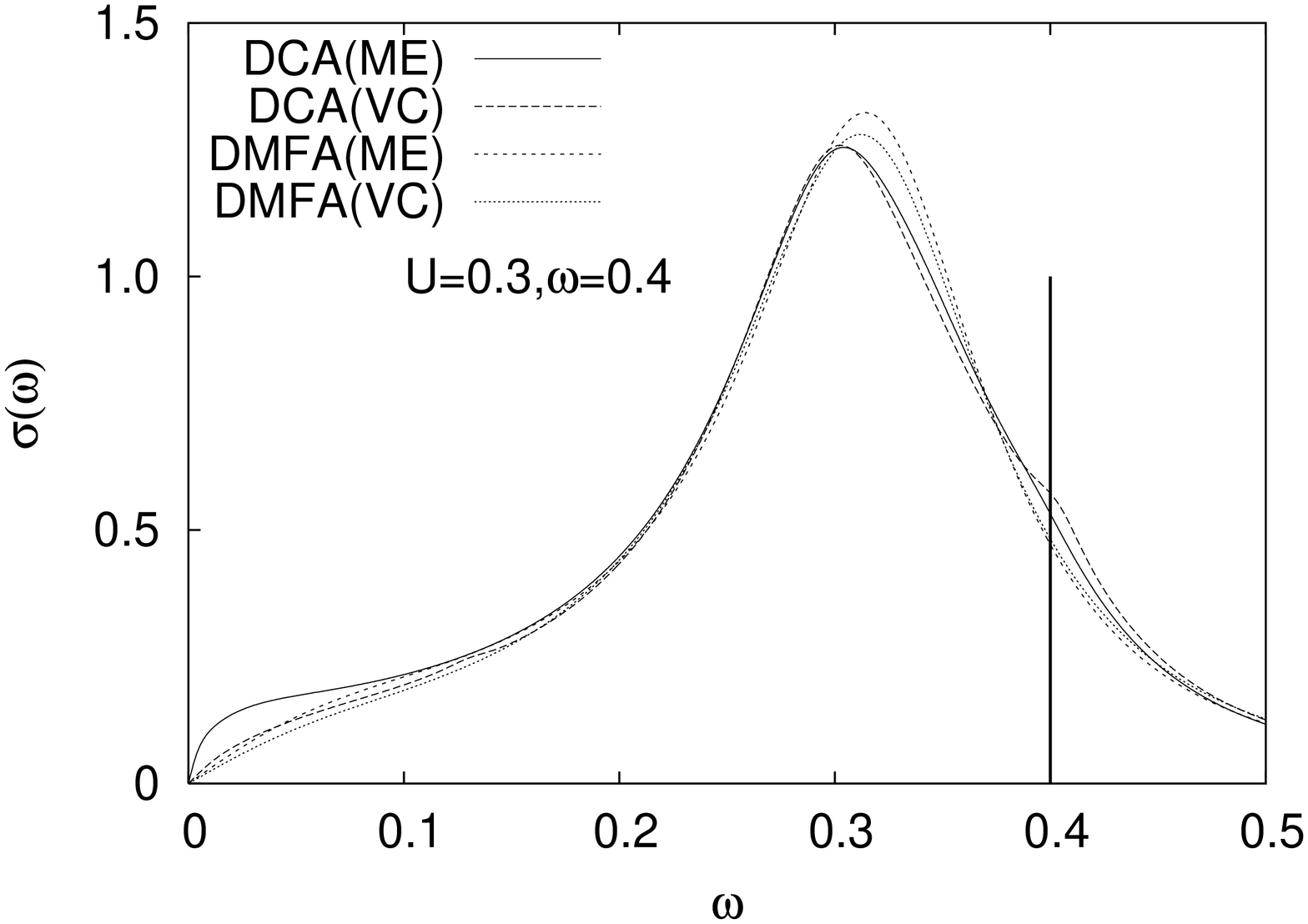}}}
\end{indented}
\caption{The renormalised phonon DOS. The upper panel gives results
for $\omega_0=0.2$, and the lower panel for $\omega_0=0.4$. All four
approximations are shown on each panel. The vertical line represents
the $\delta$-function characteristic of the non-interacting DOS in
both cases. The effect of electron phonon coupling is to shift
spectral weight to lower frequencies, and to broaden the distribution,
corresponding to a finite quasi-particle lifetime. Both panels show
similar behaviour, but the effects are more pronounced for
$\omega_0=0.4$. The DMFA results show a simple distribution with a
peak at the renormalised phonon frequency. Inclusion of vertex
corrections in the DMFA leads to a reduction in the renormalisation,
but the distribution remains similar. The inclusion of non-local
corrections to the vertex-neglected theory causes an additional shift
in the weight to very low frequencies. This shift is significant,
because it causes a large increase in the renormalised electron-phonon
coupling constant. The introduction of vertex corrections to the
non-local theory causes the distribution to return to something
similar to the original Migdal--Eliashberg result. This shows the
importance of vertex corrections in stabilising the Migdal--Eliashberg
theory.}
\label{fig:renormphondos}
\end{figure}

\begin{figure}[t]
\begin{indented}\item[]
\rotatebox{270}{\resizebox{!}{100mm}{\includegraphics{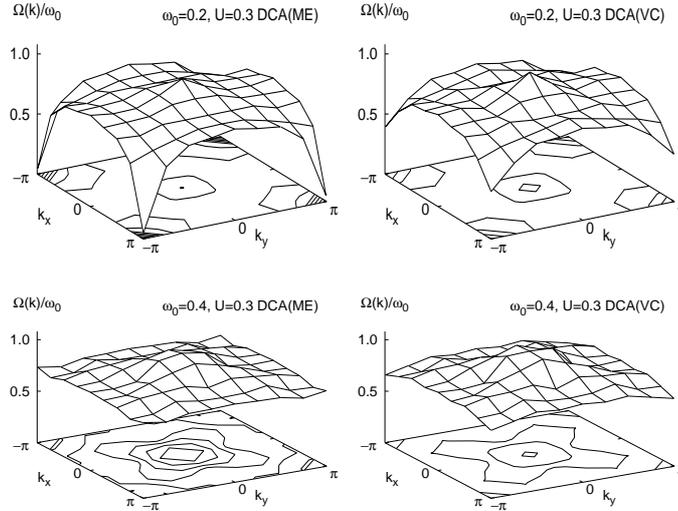}}}
\end{indented}
\caption{Renormalised phonon dispersion for the DCA. The solution from
the DMFA results in a flat dispersion, and is therefore not shown
here. The top row shows results for $\omega_0=0.2$, and the bottom row
for $\omega_0=0.4$. Vertex neglected results are shown on the left
hand side, and vertex corrected results on the right. As discussed in
the text, the vertex corrections act to reduce the renormalisation of
the phonon frequency. This can be seen to greatest effect in the
$\omega_0=0.2$ results, where the fully softened point in the
vertex-neglected theory is no longer fully softened in the vertex
corrected theory. The dispersion is relatively flat for the
$\omega_0=0.4$ case, which is reasonable since one expects a
non-renormalised dispersion as one approaches the Hubbard limit
($\omega_0\rightarrow\infty$).}
\label{fig:renormphondisp}
\end{figure}

\begin{figure}[t]
\begin{indented}\item[]
\rotatebox{270}{\resizebox{35mm}{50mm}{\includegraphics{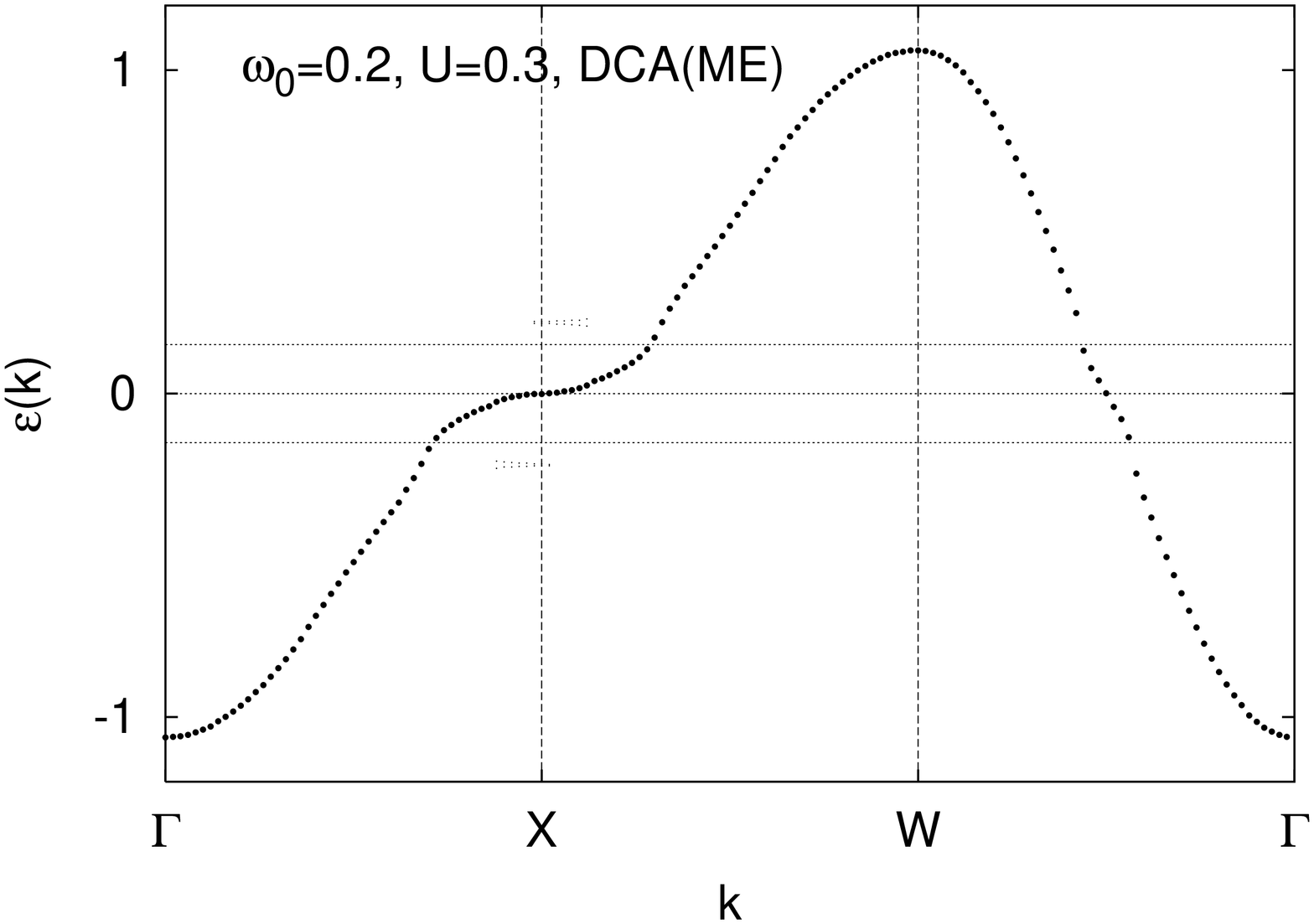}}}
\rotatebox{270}{\resizebox{35mm}{50mm}{\includegraphics{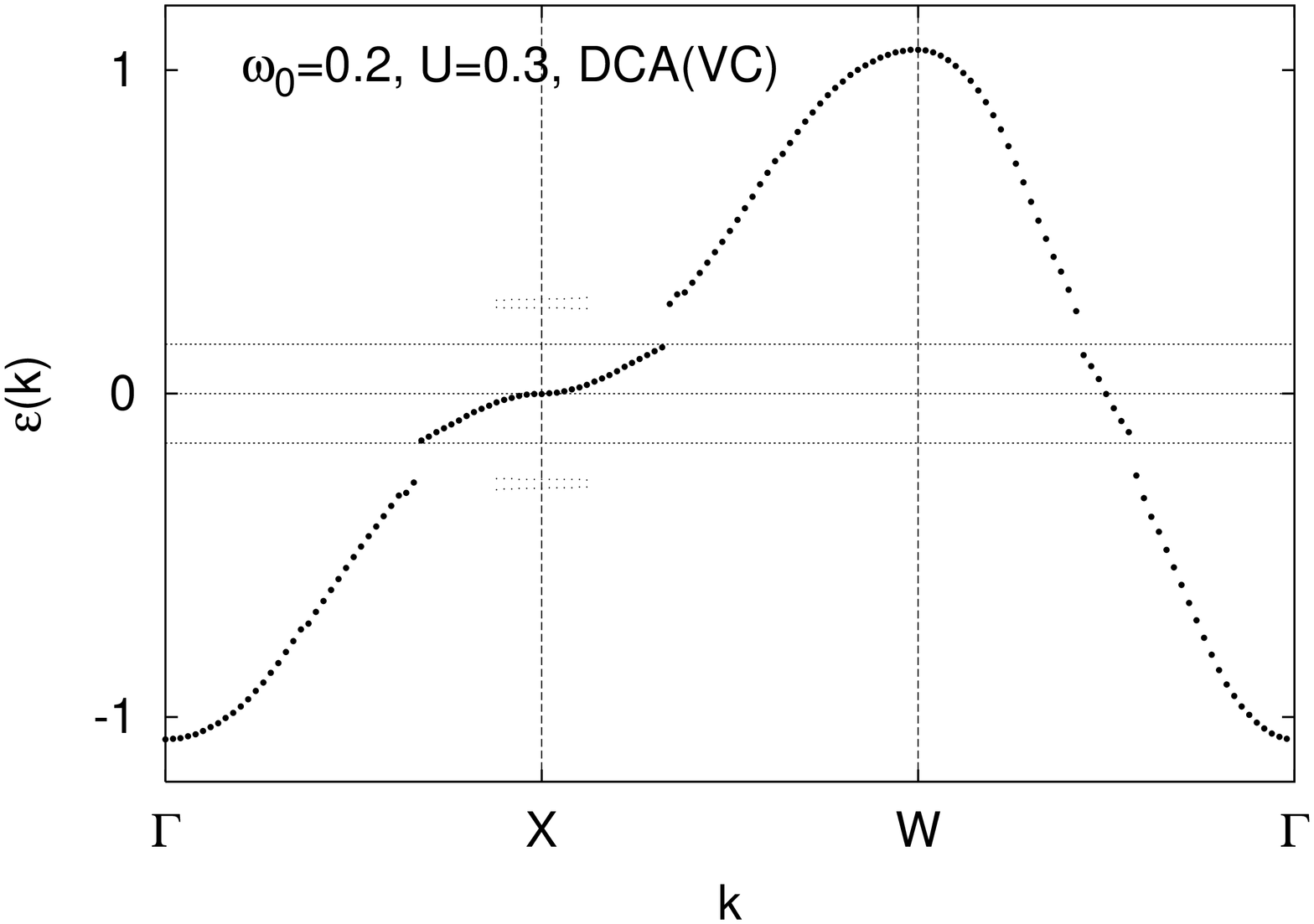}}}
\rotatebox{270}{\resizebox{35mm}{50mm}{\includegraphics{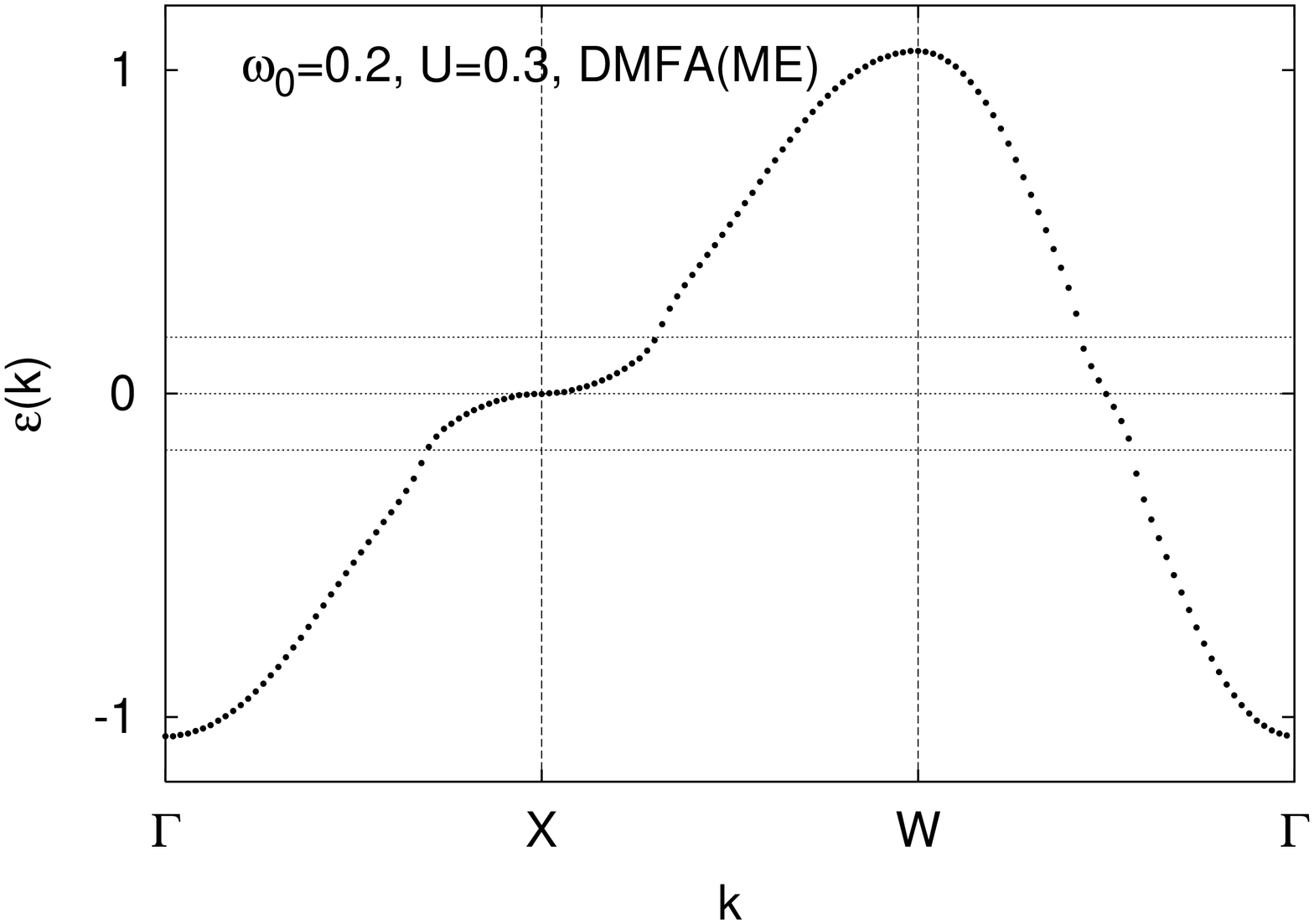}}}
\rotatebox{270}{\resizebox{35mm}{50mm}{\includegraphics{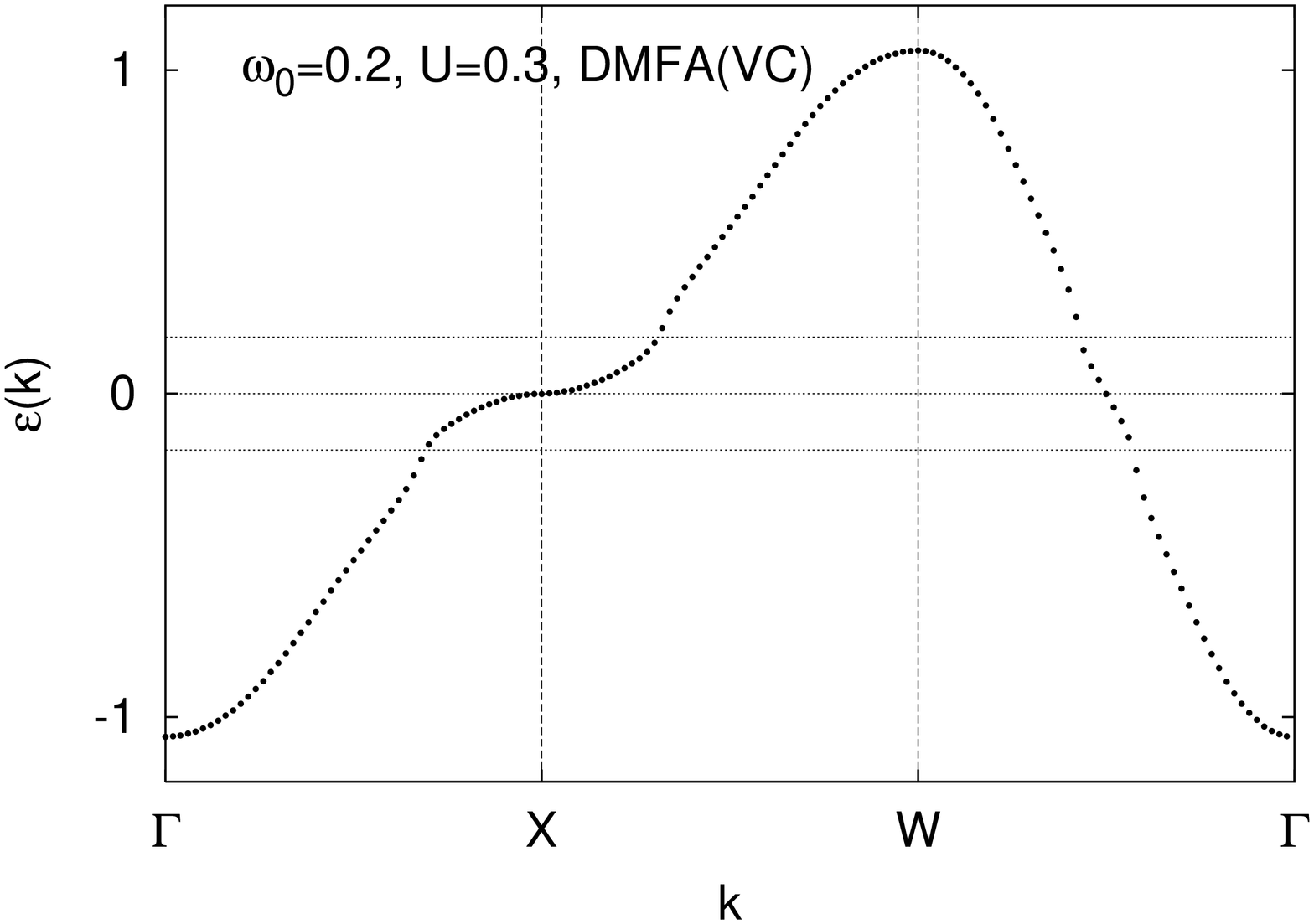}}}
\end{indented}
\caption{Electron dispersion of the Holstein model, shown for $U=0.3$,
$\omega_0=0.2$. Calculations for the DMFA and DCA with and without
vertex corrections are shown. Note how the vertex corrections
stabilise the kink. Some spurious dispersion is seen, and can be
confirmed by looking at the spectral functions in figure
(\ref{fig:arpes1}). However the weighting of these points is small,
and they are marked with small dots for completeness. All four
approximations show a kink at the renormalised phonon energy at the
($\pi,0$) point (the horizontal dotted line). The kink is especially
well defined along the $W$-$\Gamma$ line, where the 4 approaches agree
to reasonable accuracy. It is possible that the kink is slightly below
the phonon frequency for the DMFA calculations. A slight discrepancy
is reasonable, since the renormalisation of the phonon frequency in
DMFA is constant across the Brillouin zone. The kink is not well
defined along the $\Gamma$-$X$-$W$ direction, except for the DCA(VC)
calculations, where a sudden discontinuity is seen. The point of this
discontinuity does not coincide with a boundary to the coarse grained
cell, so it is expected that it is not spurious.}
\label{fig:electrondispersion1}
\end{figure}

\begin{figure}[t]
\begin{indented}\item[]
\rotatebox{270}{\resizebox{35mm}{50mm}{\includegraphics{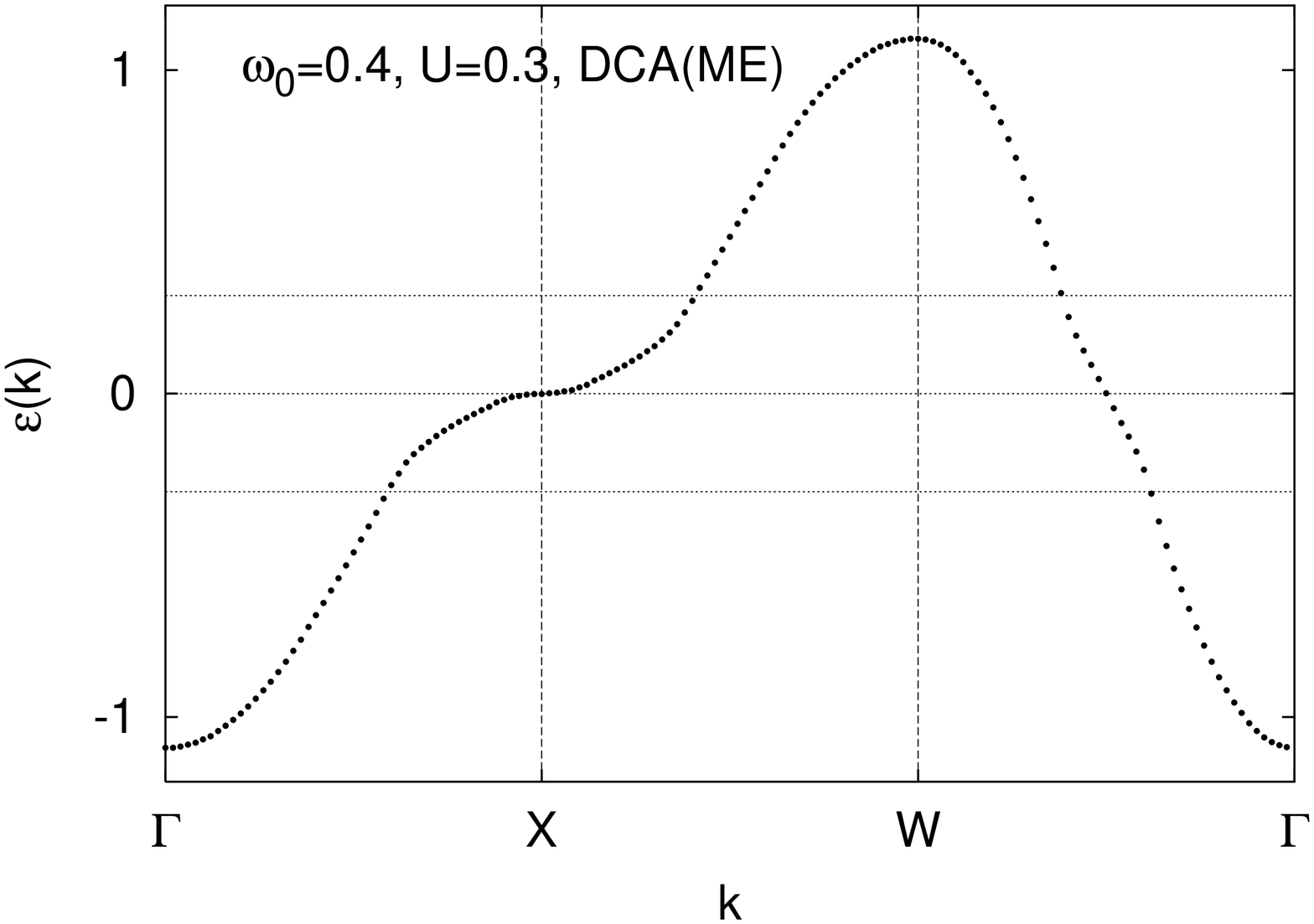}}}
\rotatebox{270}{\resizebox{35mm}{50mm}{\includegraphics{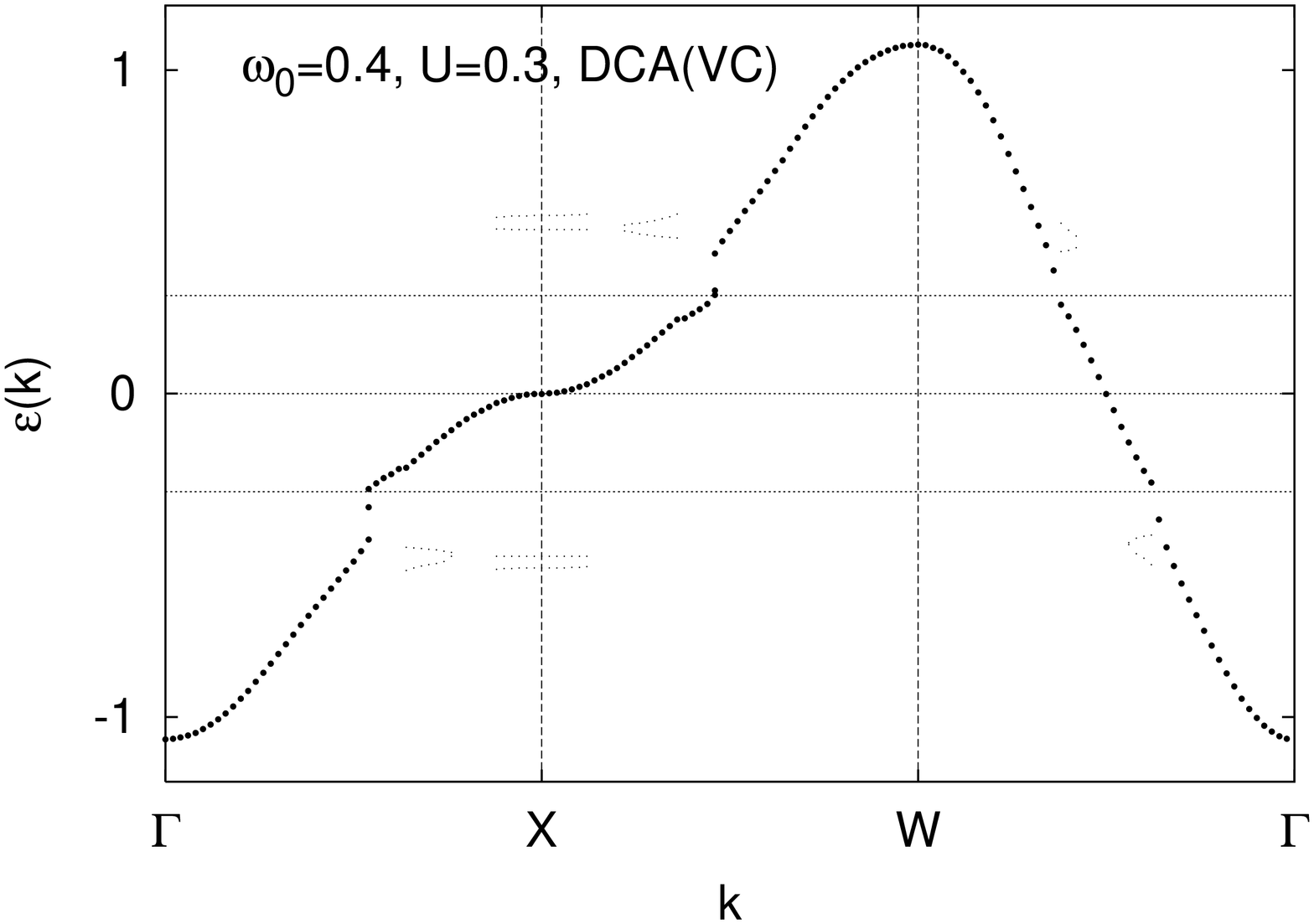}}}
\rotatebox{270}{\resizebox{35mm}{50mm}{\includegraphics{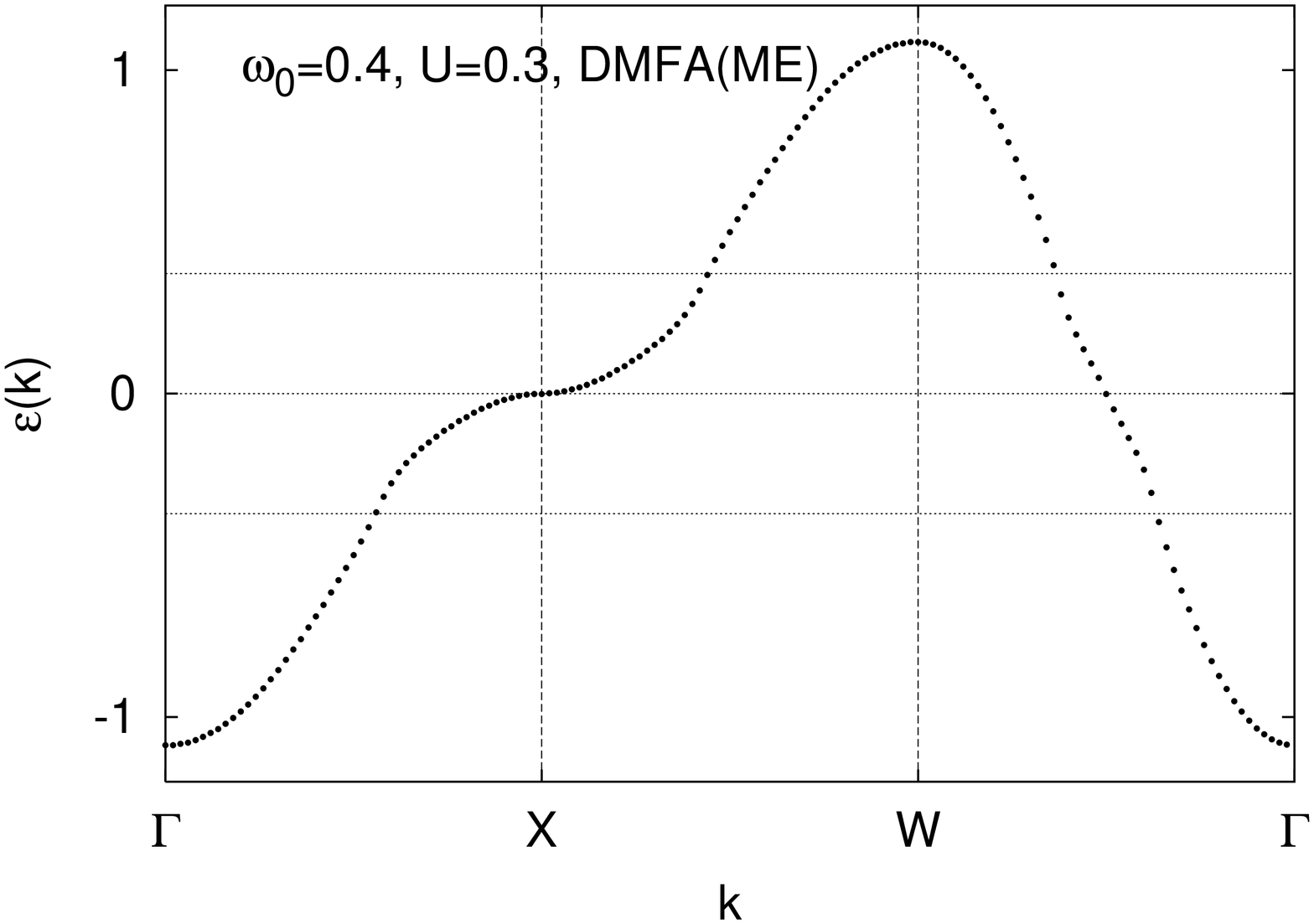}}}
\rotatebox{270}{\resizebox{35mm}{50mm}{\includegraphics{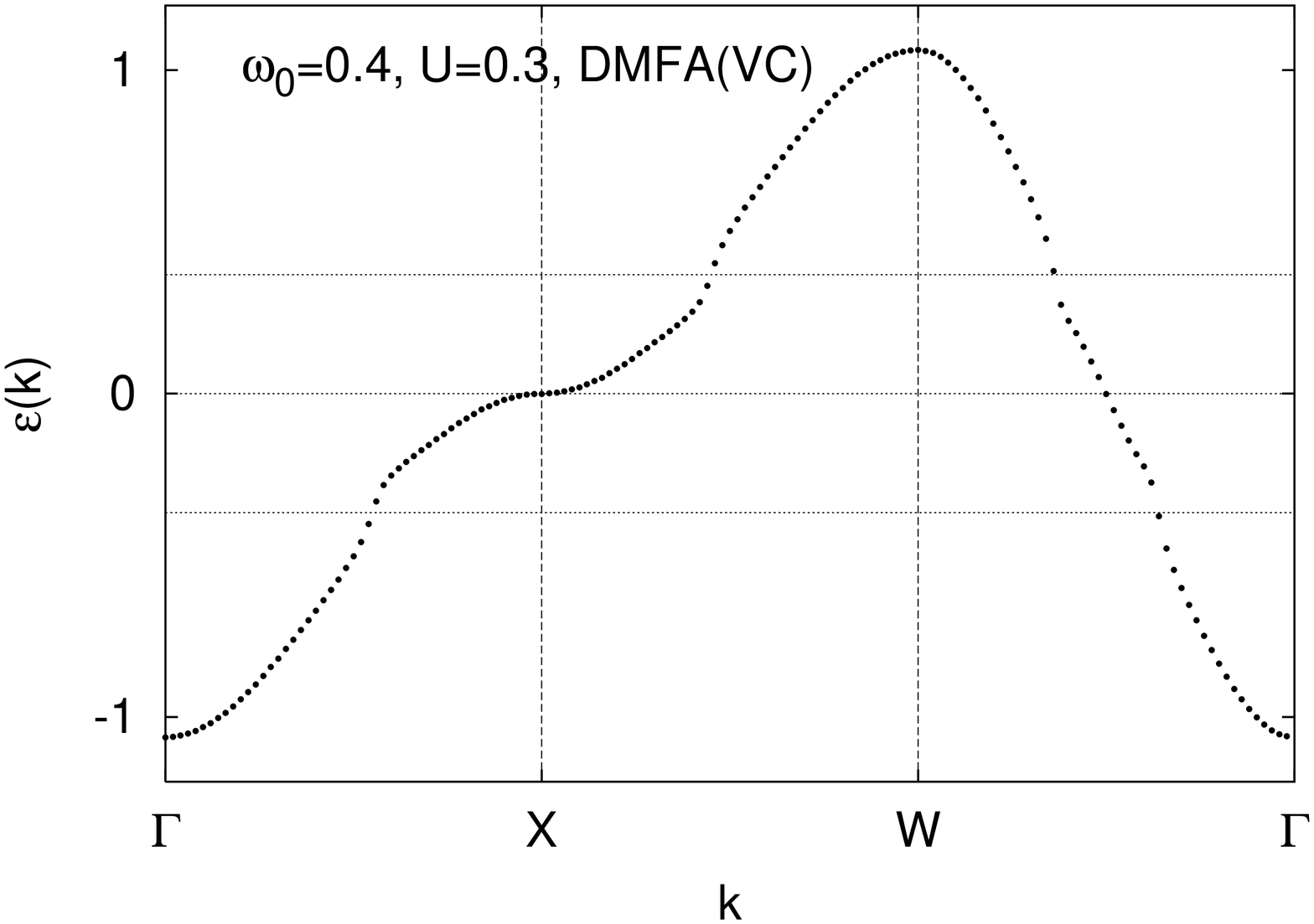}}}
\end{indented}
\caption{As figure \ref{fig:electrondispersion1} but with $U=0.3$,
$\omega_0=0.4$. As before, vertex corrections stabilise the kink. Some
spurious dispersion is seen, and can be confirmed by looking at the
spectral functions in figure (\ref{fig:arpes1}). The weighting of the
spurious points is small, but they are marked with small points for
completeness. Note that the kink energy is significantly lower than
the appropriate renormalised phonon frequency at the ($\pi,0$) point
for both of the DMFA calculations. The DCA(ME) dispersion bends
smoothly (rather than showing a sudden kink), at an energy which is
slightly lower than the phonon frequency. The DCA(VC) has kink
behaviour along the $W$-$\Gamma$ line in contrast to the other 3
approximations. Discontinuity in the dispersion along the
$\Gamma$-$X$-$W$ line can also be seen at this frequency.}
\label{fig:electrondispersion2}
\end{figure}

\begin{figure}[t]
\begin{indented}\item[]
\rotatebox{270}{\resizebox{!}{100mm}{\includegraphics{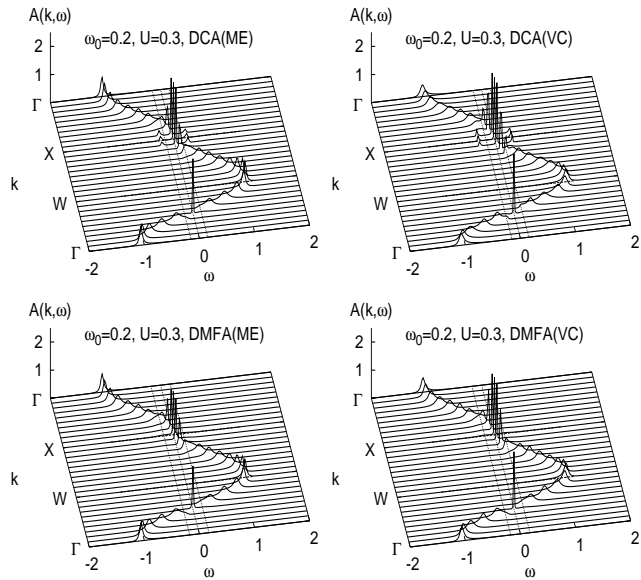}}}
\end{indented}
\caption{Spectral functions along the high symmetry directions for
$U=0.3$, $\omega_0=0.2$. Sharply peaked points close to the
Fermi-energy ($\omega=0$) show well defined quasi-particles. These
particles are suddenly damped at the renormalised phonon energy of the
$(\pi,0)$ point (parallel dotted lines). However, at higher
frequencies ($\Gamma$ and $W$ points), the damping is diminished in
the vertex-neglected case. The vertex corrections predict that damping
is maintained, even up to the largest energies. It is likely that the
DMFA overestimates the phonon energy at $(\pi,0)$. The full DCA with
vertex corrections is necessary to obtain a very crisp change at the
phonon frequency.}
\label{fig:arpes1}
\end{figure}

\begin{figure}[t]
\begin{indented}\item[]
\rotatebox{270}{\resizebox{!}{100mm}{\includegraphics{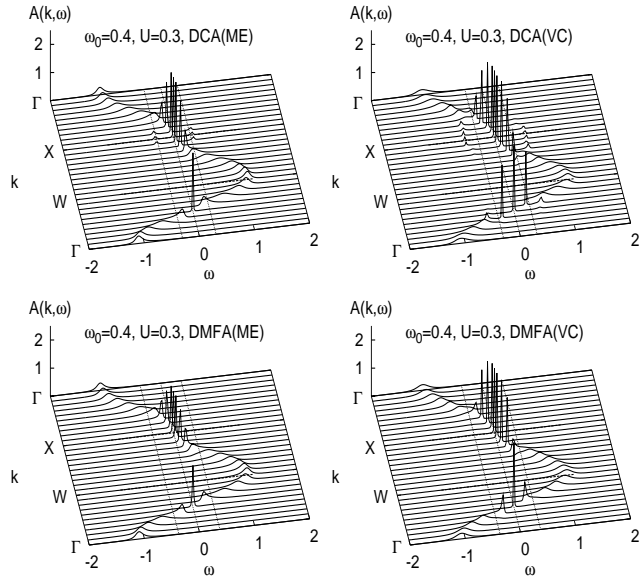}}}
\end{indented}
\caption{As figure \ref{fig:arpes1}, but with $U=0.3$,
$\omega_0=0.4$. In the case of the vertex corrected DCA, the
quasi-particles are suddenly damped at the $(\pi,0)$ phonon energy
(shown by the parallel dotted lines). This sudden increase of damping
is consistent with the experimental results in
\cite{lanzara2001a}. The vertex neglected theory consistently
estimates the energy for the damping to begin at frequencies lower
than the phonon frequency at the $X$ point. As with figure
(\ref{fig:arpes1}), the vertex corrections maintain the damping up to
the highest energies.}
\label{fig:arpes2}
\end{figure}

\begin{figure}[t]
\begin{indented}\item[]
\rotatebox{270}{\resizebox{!}{70mm}{\includegraphics{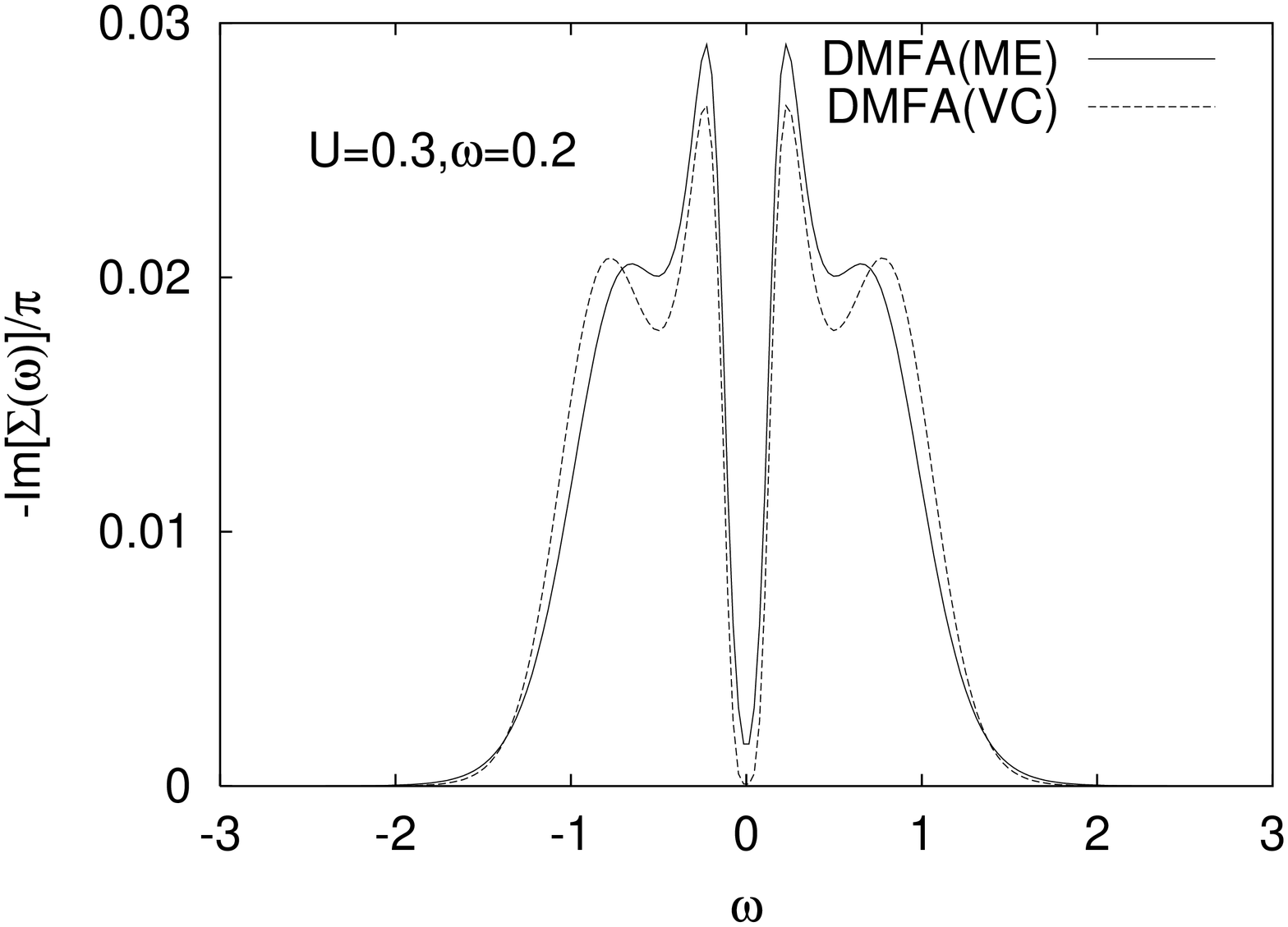}}}
\rotatebox{270}{\resizebox{!}{70mm}{\includegraphics{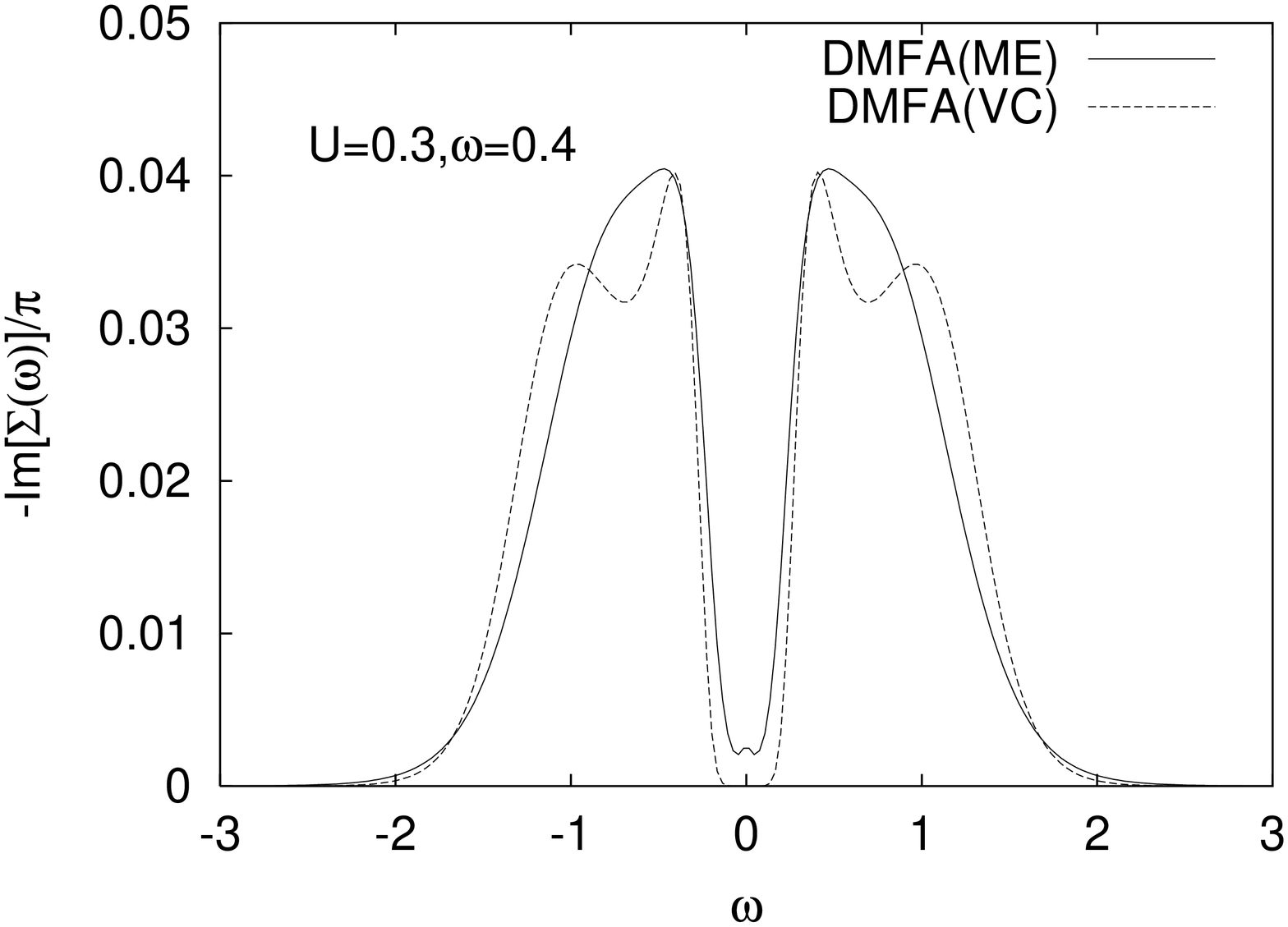}}}
\end{indented}
\caption{Self energy of the Holstein model approximated using the
DMFA, with and without vertex corrections. Since the self-energy of
the DCA varies in a complicated manner throughout the Brillouin zone,
it is not shown here. As the vertex corrections act against the Fock
term, and have a much lower energy scale, the quasi-particle damping
at the Fermi energy is suppressed. However, the zeroth order moment is
uniquely defined by the Fock term, and so the total normalisation must
be conserved. This means that weight is shifted toward higher
energies, leading to the suppression of quasi-particle states at the
top of the dispersion.}
\label{fig:selfenergy}
\end{figure}

In this section, the phonon DOS and renormalised phonon frequency are
calculated. A moderate coupling of $U=0.3$ was chosen, corresponding to
approximately one third of the band width $W=1$ (i.e. $U$ is slightly larger
than $t=0.25$). Two phonon frequencies of $\omega_0=0.2$ and $\omega_0=0.4$
were chosen to demonstrate behaviour in the region where Migdal's theorem is
(or ought to be) applicable, and behaviour in the region where vertex
corrections are important. Computations were carried out with $N_C=1$ and
$N_C=64$ in order to make comparisons between the DMFA and DCA
result, with an additional computation for $N_C=16$ in order to check
convergence in cluster size.  Although the results are not shown here, the
$N_C=16$ calculation was already close to convergence. A
cutoff of 200 Matsubara points was used at a temperature of $T=0.02$. The
vertex neglected theory was calculated first, by iterating until convergence
of 1 part in $10^{5}$. The vertex corrections were then switched on, and
iteration was continued until the required accuracy was reached.

The total (momentum integrated) phonon density of states
($\sigma(\omega)=\mathrm{Im}[D(\omega)]/\pi$) may be seen in figure
\ref{fig:renormphondos}. Since it is difficult to choose a default model for
the maximum entropy analytic continuation of Bosonic propagators
\cite{jarrell1996a}, the analytic continuation in this subsection was carried
out using the Pad\'{e} approximant method of Vidberg and Serene
\cite{vidberg1977a}. I also computed some results on the real axis for the
vertex-neglected theory, and found no significant differences. 

The top panel in figure \ref{fig:renormphondos} shows results for
$\omega_0=0.2$, and the bottom panel for $\omega_0=0.4$. On each
panel, results for all the approximations are shown to facilitate
comparison (as is the case for most graphs in this section, exceptions
will be noted with reasons). As is well documented for
Migdal--Eliashberg theory, the effect of electron-phonon coupling is
to shift spectral weight to lower frequencies, with the effect that
the phonon frequency is renormalised to a smaller value (i.e. the
phonon mode softens). By way of comparison, the thick vertical line
represents the $\delta$=function characteristic of the bare phonon
DOS.

The inclusion of vertex corrections into the DMFA calculation reduces the
softening effect. This can be seen in both panels. The reduction in
renormalisation occurs because the sign of $\Pi_{\mathrm{VC}}$ is opposite to
that of $\Pi_{\mathrm{ME}}$, i.e. the effect of particle exchange is to reduce
correlation effects.

The consequences of non-local corrections to the vertex-neglected
theory are also shown in the figure. The effects are significant,
since a proportion of the phonon spectral weight is shifted to very
low energies. As the phonon frequency is increased, the relative size
of the missing vertex corrections also increases. Therefore, the
number of low frequency states rises. This very low energy behaviour
is extremely important, since the small parameter for the
Migdal--Eliashberg perturbation theory is normally written as
$\lambda=2 g^2\int_0^\infty \sigma(\omega)\, d\omega/\omega$
\cite{carbotte1990a}. $\lambda$ is therefore significantly enhanced by
the low energy modes, especially since the DOS at low energy has
negative curvature. Examination of the partial phonon DOS for each
coarse graining cell quickly shows that the main culprit is the
$(\pi,\pi)$ cell, which is fully softened.

Vertex corrections to the non-local theory significantly reduce the
renormalisation of the $(\pi,\pi)$ point, causing the DOS to appear similar to
that calculated from the local approximation. One expects this result, since
the softening of the $(\pi,\pi)$ mode increases the total magnitude of the vertex
corrections within an iteration (by increasing the coupling constant
$\lambda$) reducing $\lambda$ over subsequent iterations. The phonon DOS finds
an equilibrium between the softening of the vertex-neglected term and the
effects of vertex corrections. This is consistent with experiments, where no
full softening of the phonon frequency at the $(\pi,\pi)$ point is observed for
systems of optic phonons \cite{mcqueeny1999a,mook1992a}. The effects of higher
order vertex corrections are expected to be consistently smaller and smaller
for this parameter range, since the lowest order vertex corrections were
sufficient to stabilise $\lambda$ to a small value. Such an effect would
presumably be irrelevant to systems of acoustic phonons, where the zone-centre
mode is already zero.

In order to elucidate the deficiencies in the vertex-neglected theory
more completely, the renormalised phonon frequency in a coarse grained
cell, $\Omega(\mathbf{K})$, may be calculated by solving the equation,
\begin{equation}
\omega_0^2-\omega^2-\omega_0^2\mathrm{Re}[\Pi(\omega,\mathbf{K})]=0
\end{equation}
In figure \ref{fig:renormphondisp}, surface plots showing the renormalised
phonon dispersion across the whole Brillouin zone are presented. Contours are
plotted beneath the dispersion to highlight the resulting symmetry. Since the
DMFA and bare phonon dispersion in the model are momentum independent, the
renormalised phonon dispersion relating to the DMFA calculation of the
Holstein model is constant across the Brillouin zone, and is therefore not
shown. In a more realistic model, the dynamical effects of the coupling
between nuclei would have to be taken into account, and one would start with a
momentum dependent dispersion \cite{ashcroft1975}. In order to apply DCA to
such a problem, one would have to apply a coarse graining procedure to the
phonon self-energy, in a similar spirit to \cite{motome2000a}. In case a
realistic model induces a momentum dependent coupling constant, then
additional coarse graining could be carried out as in
\cite{hettler2000a}. This would lead to a smooth dispersion as in the
electron case. It is clear that the complete softening of the $(\pi,\pi)$ mode
that is seen in the vertex neglected theory (top left) is absent in the vertex
corrected theory (top right). One of the main effects of non-local
fluctuations can be seen here, since the DCA predicts that electron-phonon
coupling induces a dispersion in the Holstein model (absent in the DMFA). As
shown here, the dispersion induced from the coupling of electrons to a single
Einstein mode is similar to the effects of inter-nucleus coupling in a
non-interacting system. This means that it could be quite difficult to
distinguish between the two effects in the analysis of experimental data.

\subsection{Electron dispersions and spectral functions}
\label{sec:electrondispersion}

The quasi-particle dispersion and lifetime are important physical quantities that can be measured by the ARPES
technique. The
renormalised dispersion can be calculated by solving the equation,
\begin{equation}
\omega+\mu-\epsilon_{\mathbf{k}}-\mathrm{Re}[\Sigma(\omega,\mathbf{K})]=0
\end{equation}
in analogy to the calculation of the phonon dispersion. Unlike the
phonon calculations, the electron dispersion can be determined at
all $\mathbf{k}$ points, owing to the many to one mapping involved in
coarse graining the self-energy. Here $\mathbf{K}$ is the momentum of the
coarse grained cell containing the momentum $\mathbf{k}$.

Figure \ref{fig:electrondispersion1} shows the electron dispersion of the
Holstein model, for $U=0.3$, $\omega_0=0.2$. The analytic continuation of the
self energy was carried out using the maximum entropy method
\cite{jarrell1996a}, since the normalisation of the spectral weight is known,
and a reasonable default model can easily be constructed. The Pad\'{e}
approximant technique was not found to be very stable for the analytic
continuation of self-energies. One reason for this may be that the self-energy
has a high energy behaviour of $1/\omega$, whereas the bosonic propagators
have $1/\omega^2$ behaviour, and therefore the contribution to the approximant
from higher frequencies is smaller.

By examining the dispersions calculated for $\omega_0=0.2$ it can be
seen that the main feature of the electron dispersion is a ``kink''
at the energy corresponding to the phonon frequency at the $X$ point
(the phonon frequency is marked as a horizontal dotted line). This is
persistent for all the approximations. Vertex corrections stabilise
the kink, and make it sharper. The kink is especially well defined
along the $W$-$\Gamma$ line, where the 4 approaches agree to
reasonable accuracy. It is possible that the kink is located slightly
below the renormalised phonon frequency in the dispersion calculated
using the DMFA. Such a discrepancy is expected, since the
renormalisation of the phonon frequency in DMFA is constant across the
Brillouin zone, an assumption which is only true when phonon frequency
and coupling are small as pointed out in section
\ref{sec:phonondispersion}. The kink is not well defined along the
$\Gamma$-$X$-$W$ direction, unless the dispersion is calculated using
the DCA(VC) scheme, where a sudden discontinuity is seen. The location
of this discontinuity does not coincide with the boundary to a coarse
grained cell, and is therefore not a result of the DCA. I note that
some spurious dispersion is seen (small points). However, the
quasi-particle weight of these points is small, and they are only
included for completeness.

Figure \ref{fig:electrondispersion2} shows the electron dispersion
for $U=0.3$, $\omega_0=0.4$. Note that in the results from the DMFA
calculations, the energy at which the kink occurs is significantly
lower than the $X$-point phonon frequency. Results for the DCA(ME)
bend smoothly, rather than kinking suddenly, at an energy scale, which
may be slightly lower than the phonon frequency. Only the DCA(VC) has
well defined kink behaviour along the $W$-$\Gamma$ line. The
discontinuity in the dispersion along the $\Gamma$-$X$-$W$ line can
also be seen at this frequency. It is therefore quite clear that
vertex corrections are significant for this set of parameters.

In order to calculate spectral functions at general $\mathbf{k}$, the
coarse grained self energy is used in the following equation,
\begin{equation}
G(\omega,\mathbf{k})=\frac{1}{\omega+i\eta+\mu-\epsilon_\mathbf{k}-\Sigma(\omega,\mathbf{K})}
\end{equation}
where $\mathbf{K}$ is the momentum point in the coarse graining cell
which corresponds to $\mathbf{k}$.

In the calculation of the electron dispersion, some information is
lost. For example, the quasi particle lifetime, which is related to
the width of the peak in the spectral function cannot be seen. In
order to demonstrate these effects, figure \ref{fig:arpes1} shows a
waterfall plot of spectral functions along the high symmetry
directions for $U=0.3$, $\omega_0=0.2$. The dotted lines running along the $k$ axis represent the
renormalised phonon frequency at the $(\pi,0)$ point.

Sharply peaked points can be seen close to the Fermi-energy
($\omega=0$), showing well defined quasi-particles. The
quasi-particles are suddenly damped close to the renormalised phonon
frequency corresponding to the $(\pi,0)$ point, i.e. the peaks develop
a width. At higher frequencies, the damping diminishes in the vertex
neglected case and the peaks get higher and narrower. On the other
hand, vertex corrections ensure that strong damping is maintained,
even up to the largest energies. As discussed previously, it is
thought that the DMFA overestimates the renormalised phonon frequency
at $(\pi,0)$. The full DCA with vertex corrections predicts a very
crisp kink and reduction in the quasi-particle lifetime at the $X$-point phonon
frequency.

Figure \ref{fig:arpes2} also shows spectral functions but now with the
larger phonon frequency, $\omega_0=0.4$. As with the $\omega_0=0.2$
results, the vertex-corrected DCA predicts a sudden damping of the
quasi-particles at the $(\pi,0)$ phonon energy. This sudden increase
in damping connected with the kink is consistent with the
photoemmision experiment in \cite{lanzara2001a}. The vertex neglected
theory predicts a much lower energy for the damping to begin. One can
see that the kink predicted by the DMFA begins at a lower energy that
the phonon frequency, and (for the vertex corrected DMFA) that the
damping changes smoothly, rather than suddenly before the phonon
energy, although full damping is not seen until above the phonon
energy.

The origin of the damping, and of the strong quasi-particles close to
the Fermi-surface can be demonstrated by examining the self
energy. Figure (\ref{fig:selfenergy}) shows the self energy of the
Holstein model approximated using the DMFA, with and without vertex
corrections. The self-energy of the DCA is not shown, since it varies throughout the Brillouin zone. The
form of the DCA self-energies are similar. As the vertex corrections act to reduce the self-energy
at low energy scales, the self energy at the Fermi energy is
suppressed, leading to well defined quasi-particles at low
energies. However, the high frequency behaviour is defined by the
first-order term in the self energy, and this can be related to the
normalisation of the self-energy spectrum. Therefore the total
normalisation is seen to be conserved, regardless of the approximation
used. If the weight of the self-energy is reduced at low frequencies,
this weight must be shifted toward higher energies,
leading to the suppression of quasi-particle states at the top of the
dispersion.

\section{Summary}
\label{section:summary}

In this paper, I applied the newly developed dynamical cluster
approximation to the calculation of the electron dispersion, induced
phonon dispersion, spectral functions and phonon DOS for the Holstein
model. A theory neglecting vertex corrections (related to
Migdal--Eliashberg theory) and another incorporating the lowest order
vertex correction were investigated. In contrast to previous work,
non-local corrections were included. The application of DCA to this
problem was crucial, since the double three-fold (four-fold including
Matsubara summations) integration involved with the vertex corrected
self-energy of phonons and electrons means that only small clusters
can be considered, precluding finite size calculations.

The effect of non-local corrections to the vertex neglected theory of
electron-phonon interactions is dramatic. Inclusion of non-local
corrections to the theory results in a complete softening of the
phonon modes at and close to the $(\pi,\pi)$ point, which increases the
effective coupling, $\lambda$. Therefore, a first-principles approach
to the electron-phonon problem in two dimensions which neglects the
effects of vertex corrections, but keeps the effects of non-local
corrections, can only be applied to a very small range of phonon
frequencies and couplings.

Vertex corrections act against the non-vertex corrected theory in both
the phonon and electron self-energies. In the case of the phonon
self-energy, which renormalises the phonon frequency, the correction
acts to avert the complete softening of the mode. Therefore, $\lambda$
is stabilised. The theory with vertex corrections also acts to
stabilise the kink in the electron dispersion at large phonon
frequencies, and causes an enhancement in the kink.

The results of this study have wide ranging implications for the
application of Migdal--Eliashberg theory to the electron-phonon
problem. Dispersion can be induced in the phonon modes, making it
difficult to estimate the true coupling from fits to experimental
data. Vertex corrections act to return the system to the Migdal
parameter regime. Therefore, it is not surprising that the
Migdal--Eliashberg theory has been so successful in describing the
properties of a wide range of superconductors.

There is a large scope for extension to this work. The most direct and
straightforward extensions would be to investigate the superconducting
phase, the phase diagram and the 3D problem. It would also be
interesting to study the effects of even higher-order vertex
corrections. This is not practical within perturbation theory, since
an implementation would involve would involve many 3-fold
integrations. An alternative method for including vertex corrections
would be to employ a quantum Monte-Carlo simulation. Such work is
currently in progress.

\ack

JPH would like to acknowledge support from the Guest Scientist program
of the Max-Planck-Institute f\"{u}r Physik Komplexer Systeme, and
useful discussions with Mark Jarrell, Jim Freericks, Nick
d'Ambrumenil and Emma Chung.

\vspace{10mm}
\bibliographystyle{unsrt}
\bibliography{dcaholstein}

\end{document}